\documentclass[aps,prl,twocolumn,amsmath,amssymb,footinbib,showpacs,longbibliography,superscriptaddress
]{revtex4-1}

\newcommand{\Jnature}{Nature (London)}
\newcommand{\Jnatphys}{Nat. Phys.}
\newcommand{\Jnatphot}{Nat. Photonics}
\newcommand{\Jnatcomm}{Nat. Comm.}

\newcommand{\JSciRep}{Sci. Rep.}

\newcommand{\Jscience}{Science}

\newcommand{\Jpnas}{Proc. Nat. Acad.  Sci.}

\newcommand{\Jprx}{Phys. Rev. X}
\newcommand{\Jprl}{Phys. Rev. Lett.}

\newcommand{\Jpra}{Phys. Rev. A}
\newcommand{\Jprb}{Phys. Rev. B}

\newcommand{\Jrmp}{Rev. Mod. Phys.}

\newcommand{\Jepl}{Europhys. Lett.}

\newcommand{\Jjetp}{Sov. Phys. JETP}

\newcommand{\JRepProgPhys}{Rep. Prog. Phys.}

\newcommand{\JjphysA}{J. Phys. A: Math. Theor.}
\newcommand{\JjphysB}{J. Phys. B: At. Mol. Opt. Phys.}
\newcommand{\JjphysC}{J. Phys. C: Solid State Phys.}

\newcommand{\Jadvphys}{Adv. Phys.}

\usepackage[english]{babel}
\usepackage{latexsym}
\usepackage{graphics}
\usepackage{subfigure}
\usepackage{epsfig}
\usepackage{color}
\usepackage{hyperref}
\usepackage{braket} 
\usepackage[T1]{fontenc}
\usepackage[latin9]{inputenc}
\usepackage{amstext}
\usepackage{amssymb}
\usepackage{graphicx}
\usepackage{esint}
\usepackage{braket}
\usepackage{babel}
\usepackage{amsmath}
\usepackage{siunitx}

\hypersetup{
colorlinks=true,
citecolor=blue,
linkcolor=red,
urlcolor=black
}


\newcommand{\e}{\textrm{e}}

\newcommand{\Er}{E_{\textrm{r}}}

\newcommand{\kB}{k_{\textrm{\tiny B}}}
\newcommand{\fs}{f_{\textrm{s}}}
\newcommand{\ns}{n_{\textrm{s}}}

\newcommand{\vect}[1]{{\mathbf{#1}}}
\newcommand{\rr}{\mathbf{r}}

\newcommand{\atwod}{a_{\textrm{\tiny 2D}}}

\newcommand{\tildeg}{\tilde{g}}
\newcommand{\tildegnot}{\tildeg_0}

\newcommand{\lambdaT}{\lambda_\textrm{T}}

\newcommand{\muc}{\mu_\textrm{c}}
\newcommand{\nc}{n_\textrm{c}}

\newcommand{\lettersection}[1]{\section{#1}}
\renewcommand{\lettersection}[1]{\paragraph*{#1.---}}

\begin{document}

\title{
Thermodynamic Phase Diagram of Two-Dimensional Bosons in a Quasicrystal Potential
}

\author{Zhaoxuan Zhu}
\affiliation{CPHT, CNRS, Ecole Polytechnique, IP Paris, F-91128 Palaiseau, France}

\author{Hepeng Yao}
\affiliation{Department of Quantum Matter Physics, University of Geneva, 24 Quai Ernest-Ansermet, CH-1211 Geneva, Switzerland}

\author{Laurent Sanchez-Palencia}
\affiliation{CPHT, CNRS, Ecole Polytechnique, IP Paris, F-91128 Palaiseau, France}

\date{\today}

\begin{abstract}
Quantum simulation of quasicrystals in synthetic bosonic matter now paves the way to the exploration of these intriguing systems in wide parameter ranges.
Yet thermal fluctuations in such systems compete with quantum coherence, and significantly affect the zero-temperature quantum phases.
Here we determine the thermodynamic phase diagram of interacting bosons in a two-dimensional, homogeneous quasicrystal potential.
Our results are found using quantum Monte Carlo simulations. Finite-size effects are carefully taken into account
and the quantum phases are systematically distinguished from thermal phases.
In particular, we demonstrate stabilization of a genuine Bose glass phase against the normal fluid in sizable parameter ranges. 
Our results for strong interactions are interpreted using a fermionization picture and experimental relevance is discussed.
\end{abstract}

\maketitle

The discovery of quasiperiodic structures in plane tilings~\cite{penrose1974} and material science~\cite{shechtman1984,levine1984} has profoundly altered our dichotomous perception of order and disorder.
Lying at the interface of the two realms, quasicrystals display a number of
intriguing properties, including
unusual localization and fractal properties,
anomalous critical scalings,
and phasonic degrees of freedom~\cite{senechal1995,steuer2004,steurer2018,edagawa2000,freedman2006,cubitt2015}.
So far, quasicrystals have been observed in their natural state in meteorites~\cite{bindi2012,bindi2015} and nuclear blast residues~\cite{bindi2021} or in the laboratory after fast solidification of certain alloys~\cite{shechtman1984,shechtman1985}, and
have been extensively studied in solid-state physics~\cite{shechtman1984,steuer2004,steurer2018,kamiya2018,ahn2018,yao2018b}.
Moreover, artificial quasicrystals can now be engineered in synthetic quantum matter with unique control knobs, using photonic crystals~\cite{chan1998,lahini2009,freedman2006,vay2013}, quantum fluids of light~\cite{tanese2014,barboux2017,goblot2020}, and ultracold quantum gases~\cite{lsp2010,modugno2010,mace2016}.
In the latter, defectless and phononfree quasicrystal potentials can be emulated in a variety of configurations
using appropriately-arranged sets of laser beams~\cite{guidoni1997,grynberg2000,lsp2005,jagannathan2013,viebahn2019,sbroscia2020}.
Furthermore, two-body interactions can be tuned using magnetic control~\cite{lewenstein2007,bloch2008,pollack2009,chin2010}, hence paving the way to the exploration of quantum phase diagrams in wide parameter ranges.

In past years, one-dimensional (1D) quasiperiodic models of ultracold atoms have been discussed quite exhaustively~\cite{roth2003,
damski2003,
roati2008,
fallani2007,
roscilde2008,
roux2008,
gadway2011,
derrico2014,
gori2016,
an2018,
an2021,
yao2019,
yao2020,
lellouch2014,
iyer2013,
schreiber2015,
khemani2017,
kohlert2019,
liu2022}
but exploration of their 2D counterparts has only recently gained momentum, mostly in tight-binding models~\cite{johnstone2019,johnstone2021,gottlob2023}.
So far, theoretical and experimental work has demonstrated the emergence of quasicrystalline order through matterwave interferometry~\cite{lsp2005,viebahn2019},
Anderson-like localization~\cite{lsp2005,szabo2020,sbroscia2020},
and Bose glass (BG) physics~\cite{johnstone2019,johnstone2021,sbroscia2020,gautier2021}.
The BG is an emblematic compressible insulator, characteristic of disordered or quasi-disordered systems and distinct from the superfluid (SF) and Mott insulator (MI) phases, which also appear in periodic systems~\cite{giamarchi1987,giamarchi1988,fisher1989}.
In bosonic models, however, thermal fluctuations compete with (quasi-)disorder, which
has so far hindered the observation of the BG phase~\cite{derrico2014,gori2016}.
It has been recently proposed that this issue may be overcome by scaling up characteristic energies using shallow quasiperiodic potentials~\cite{yao2019}.
Up to now, this has been investigated only in 1D~\cite{yao2020} and 2D harmonically trapped~\cite{ciardi2022} systems.
In contrast, the case of a 2D Bose gas with genuine long-range quasicrystal order remains unexplored.
Moreover, the central issue of discriminating the BG phase from trivial thermal phases has been hardly addressed.
As argued below, this cannot be achieved as in 1D and requires specific analysis in 2D. 

 \begin{figure*}[t!]
    \centering
    \subfigure{\includegraphics[width=0.305\textwidth]{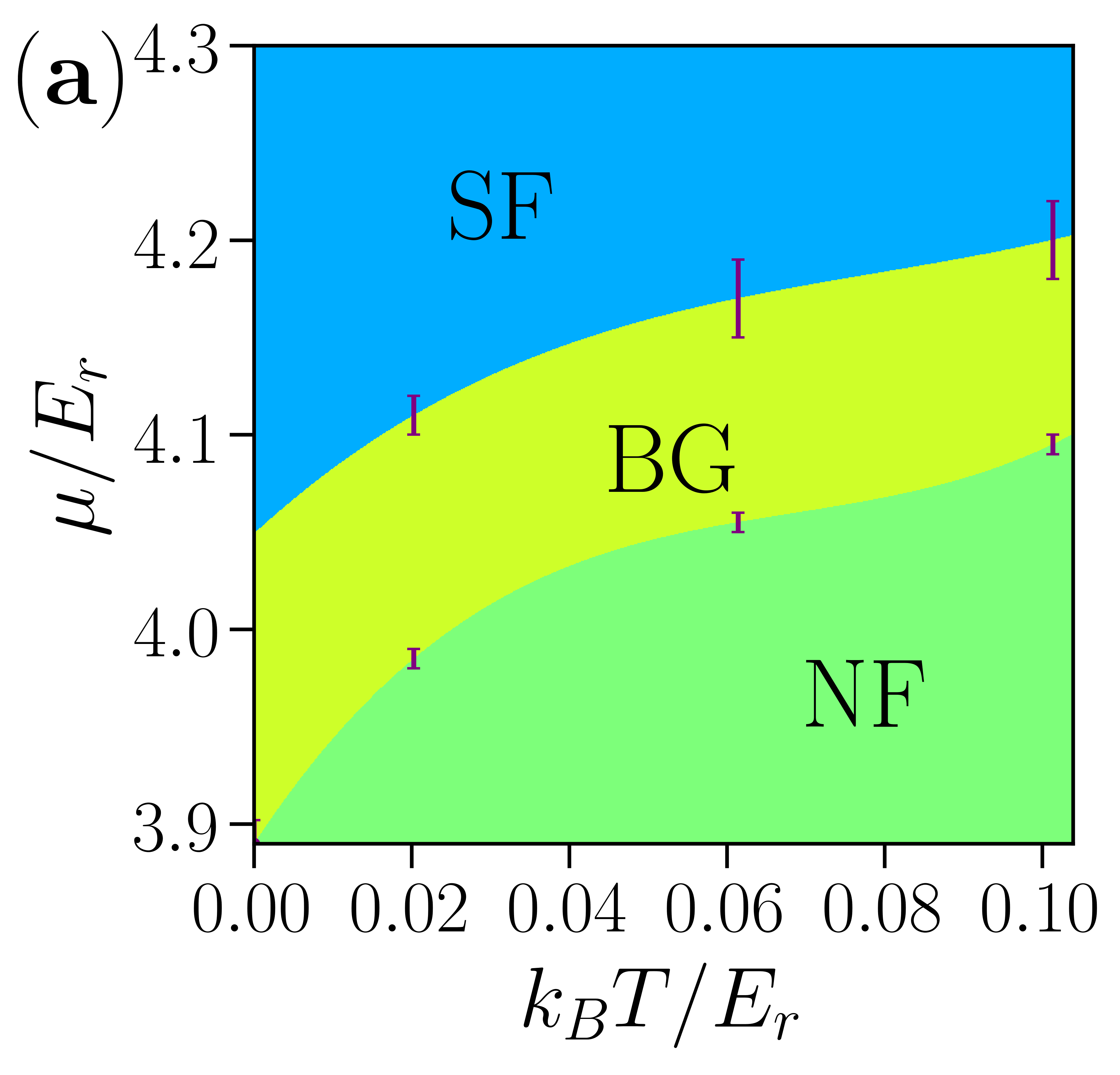}}
~~~    \subfigure{\includegraphics[width=0.301\textwidth]{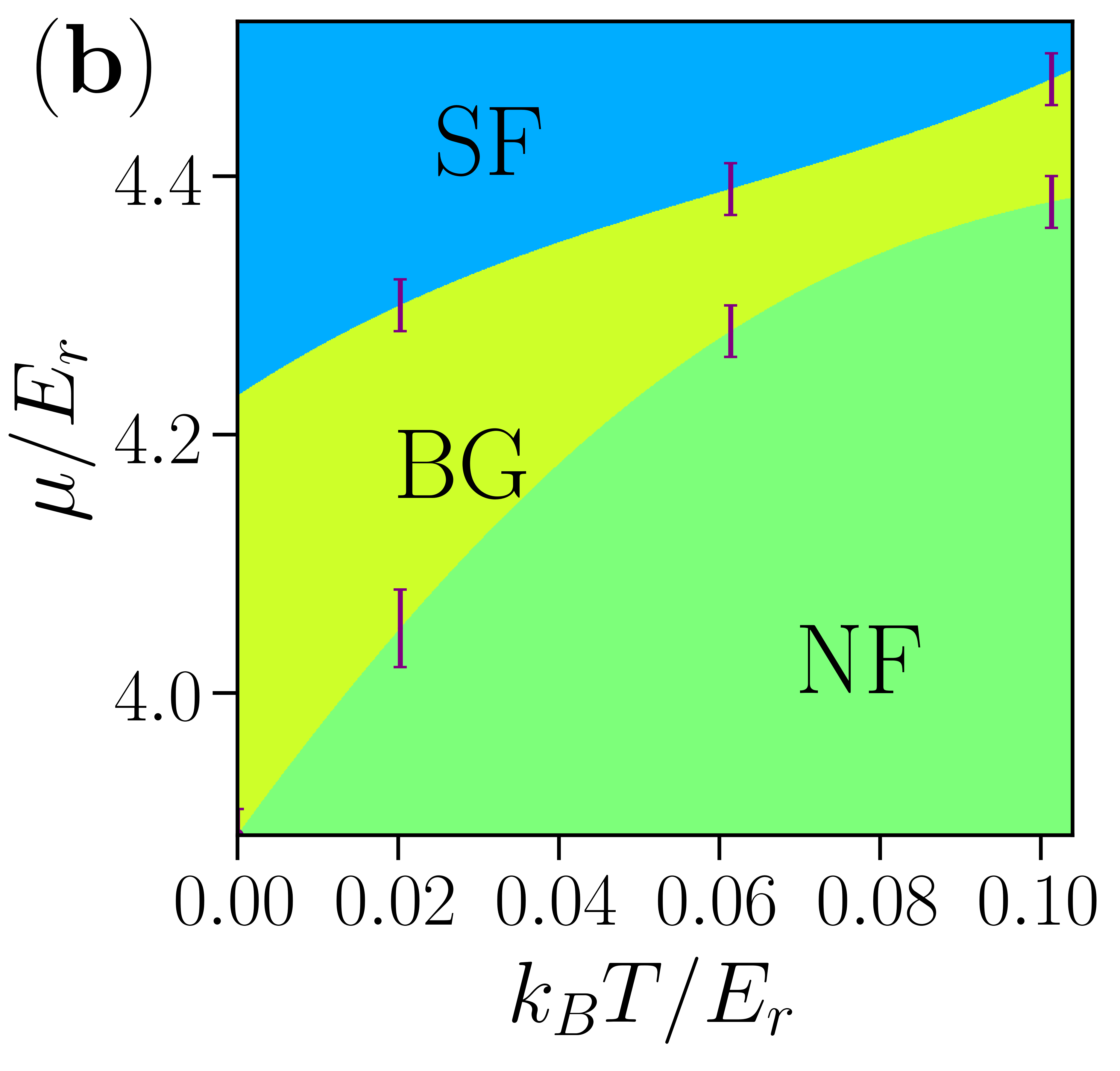}} 
~~~     \subfigure{\includegraphics[width=0.2975\textwidth]{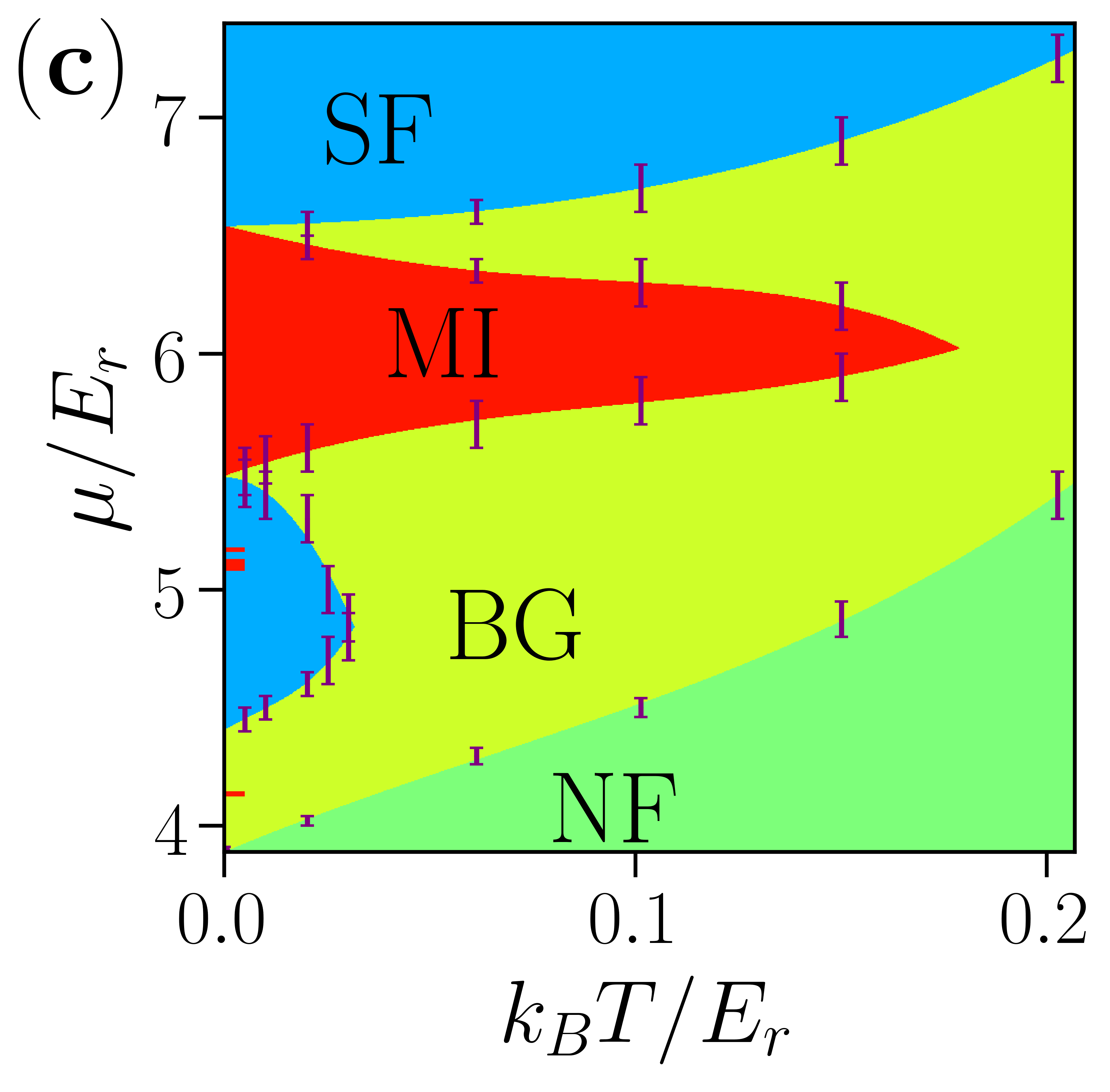}} 
\caption{\label{fig:phasediagrams}
Thermodynamic phase diagrams of 2D bosons in the eightfold quasicrystal potential of Eq.~(\ref{eq:QPpotential}) with amplitude
$V_0=2.5\Er$ and different interaction strengths,
(a)~$\tildegnot=0.05$,
(b)~$\tildegnot=0.86$,
and (c)~$\tildegnot=5$.
The quantum phases, SF (blue), BG (yellow), and MI (red), are distinguished from the NF regime (green).
Note the small MI lobes in panel~(c) at $\mu \simeq 4.1\Er$ and  $\mu \simeq 5.1\Er$, which survive only at very low temperatures.
QMC results are shown as data points with errorbars, while color boundaries are guides to the eye.
}
 \end{figure*}

In this Letter, we determine the first thermodynamic phase diagrams of weakly to strongly interacting 2D Bose gases in a shallow quasicrystal potential at  finite temperatures.
Quantum Monte Carlo simulations are performed in quasicrystal, homogeneous potentials and finite-size effects are carefully taken into account.
The SF, MI, and BG quantum phases, induced by the competition of interactions and quasicrystal potential, are systematically discriminated from the normal fluid (NF), which is instead dominated by thermal fluctuations.
Most importantly, we find that the BG phase survives up to significantly high temperatures. Our results in the strongly-interacting regime are interpreted using a fermionization picture and implications to experiments in ultracold atom systems are discussed.

\lettersection{Model}
The dynamics of the 2D Bose gas is governed by the Hamiltonian
\begin{eqnarray}\label{eq:Hamiltonian}
\hat{H} 
& = & 
\int d\rr\,
\Psi(\rr)^\dagger \left[\frac{-\hbar^2\mathbb{\nabla}^2}{2m} + V(\rr) \right]\Psi(\rr),
\\
& & + 
\frac{1}{2}\int d\rr\, d\rr'\, \Psi(\rr)^\dagger\Psi(\rr')^\dagger U(\rr-\rr')\Psi(\rr')\Psi(\rr),
\nonumber
\end{eqnarray}
where $\Psi(\rr)$ is the bosonic field operator at position $\rr$ and $m$ is the particle mass.
The quasicrystal potential,
\begin{equation}\label{eq:QPpotential}
V(\rr) = V_0 \sum_{k=1}^4 \cos^2 \left(\vect{G}_k \cdot \rr \right),
\end{equation}
is the sum of four standing waves with amplitude $V_0$ and lattice period $a=\pi/\vert\vect{G}_k\vert$,
and successively rotated by an angle of \ang{45}.
This potential is characterized by an eightfold discrete rotational symmetry, incompatible with periodic order, hence forming a quasiperiodic pattern.
The bosons interact via the two-body scattering potential $U(\rr-\rr')$.
At low energy, the collisions are dominated by s-wave scattering and hence fully characterized by the sole 2D scattering length $\atwod$.
Due to the logarithmic scaling of the interaction strength versus the scattering length in 2D~\cite{petrov2000a,petrov2001,pricoupenko2007}, it is convenient to use the interaction parameter
\begin{equation}\label{eq:coupling0}
\tildegnot = \frac{2\pi}{\ln(a/\atwod)}.
\end{equation}
The model considered here is similar to that recently emulated in ultracold-atom quantum simulators in Refs.~\cite{viebahn2019,sbroscia2020}.
The typical potential amplitude $V_0$ ranges from zero to a few tens of recoil energies, $\Er = \pi^2\hbar^2/2ma^2$. In the eightfold quasicrystal potential~(\ref{eq:QPpotential}), the critical amplitude for single-particle localization is $V_0 \simeq 1.76\Er$~\cite{gautier2021}.
So far, ultracold bosons in such 2D quasicrystal potential have been studied for vanishing or weak interactions, up to $\tildegnot \simeq 0.86$~\cite{sbroscia2020}. However, significantly higher values can be realized using transverse confinement or Feshbach resonances, up to the strongly-interacting regime, where $\tildegnot \sim 1-5$~\cite{ha2013}.
The typical temperature in ultracold atom experiments is $\kB T / \Er \sim 0.01-0.5$ with $\kB$ the Boltzmann constant.

\lettersection{Finite-temperature phase diagrams}\label{sec:diagrams}
Figure~\ref{fig:phasediagrams} shows the thermodynamic phase diagrams of the interacting Bose gas in a quasicrystal potential of amplitude $V_0=2.5\Er$ (above the critical localization potential) for three values of the interaction parameter $\tildegnot$, ranging from weak to strong interactions.
The numerical calculations are performed using path-integral quantum Monte Carlo (QMC) simulations within the grand-canonical ensemble at temperature $T$ and chemical potential $\mu$.
Details about the analysis of the numerical results, in particular as regards finite-size effects, appear below.
In brief, we compute the compressibility $\kappa = L^{-2}\partial N/\partial \mu$,
where $N$ is the average particle number and $L$ the system's linear size,
as well as the superfluid fraction $\fs$, found using the winding number estimator with periodic boundary
conditions~\cite{ceperley1995}.
These two quantities are sufficient to identify the expected zero-temperature quantum phases:
SF ($\kappa \neq 0$ and $\fs \neq 0$),
BG ($\kappa \neq 0$ and $\fs = 0$),
and MI ($\kappa = 0$ and $\fs = 0$).
For high enough temperatures, however, one expects a NF regime, dominated by thermal fluctuations. It is characterized by a finite compressibility and absence of superfluidity ($\kappa \neq 0$ and $\fs = 0$), just as the BG phase.

To discriminate a genuine BG against a trivial NF, we use the criterion that phase coherence and superfluidity must be destroyed by quasi-disorder and not thermal fluctuations~\cite{giamarchi1987,giamarchi1988}.
In 1D, any finite temperature destroys superfluidity so that the BG phase is strictly well defined only at zero temperature.
In practice, it is thus sufficient to identify a NF by the onset of a sizable temperature dependence of characteristic quantities, as done in Refs.~\cite{derrico2014,gori2016,yao2020}.
In dimensions higher than one, however, quantum phases can survive at finite temperature while showing a significant temperature dependence of the characteristic quantities, and the above criterion breaks down.
To discriminate the BG from the NF in the 2D Bose gas, we thus proceed differently and systematically compare the obtained phases in the presence of the quasicrystal potential with those of the homogeneous gas for the same temperature and the same average number of particles: If the gas is a SF in the absence of the quasicrystal potential, we identify a BG phase as soon as the quasicrystal potential amplitude is sufficient to destroy superfluidity; Otherwise, we have a NF.

\begin{figure}[t!]
         \centering
         \includegraphics[width = 0.95\columnwidth]{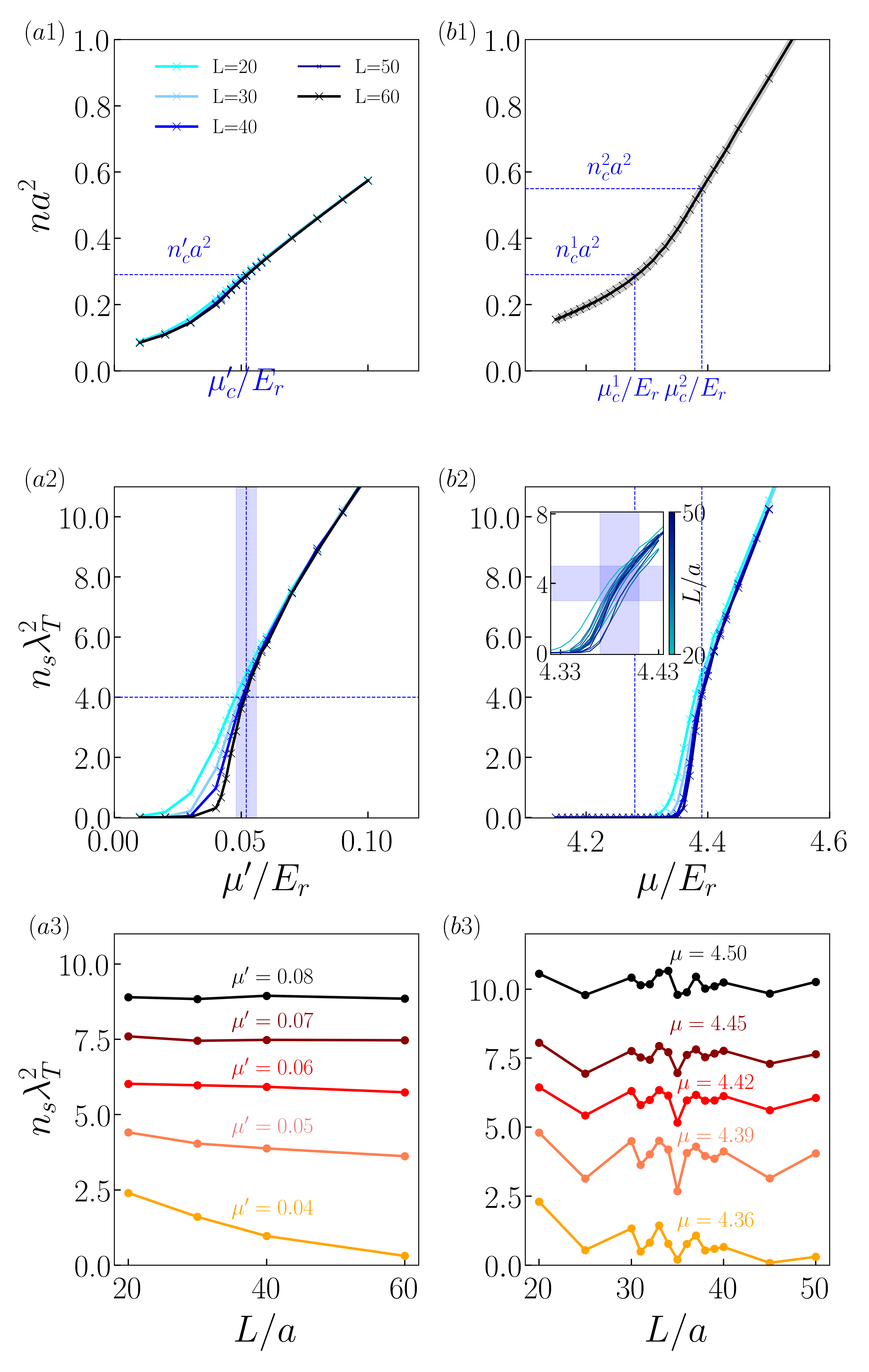}   
\caption{\label{fig:results}
Total particle density (upper row) and SF density versus chemical potential (middle row) as well as versus system size (lower row)
for a 2D Bose gas with interaction strength $\tildegnot=0.86$ and temperature $T=0.06\Er/\kB$, in the absence (left column) and in the presence (right column, $V_0=2.5\Er$) of a quasicrystal potential.
The QMC calculations are performed in square boxes for different linear sizes $L$ corresponding to the different line colors in the upper and middle rows.
The QMC statistical errorbars are smaller than the markers.
In panel~(b1), the shaded area corresponds to the standard deviation of the density fluctuations with the system size.
The Inset in~(b2) is a magnification of its main panel in the vicinity of the critical point for many system sizes with $L/a \in [20,50]$
and the shaded area is the construction to locate the SF-to-BG transition point.
Panels~(a3) and (b3) show the SF density as a function of $L$ for various chemical potentials in the vicinity of the SF transition.
}
 \end{figure}

\lettersection{Superfluid-to-Bose glass transition}\label{sec:BG}
Typical QMC results for the total particle density $n=N/L^2$ and the SF density $\ns=\fs \times n$ versus chemical potential for various system sizes are shown on Fig.~\ref{fig:results} for intermediate interaction strength and temperature, $\tildegnot=0.86$ and $T=0.06\Er/\kB$.
Similar results are found in all ranges of temperature, chemical potential, and interaction strength considered for the phase diagrams of Fig.~\ref{fig:phasediagrams}, up to the MI phase relevant for strong interactions (see below).
In the absence of a quasicrystal potential, $V_0=0$ [Fig.~\ref{fig:results}(a), left column], the QMC results show a clear NF-to-SF transition, characteristic of the expected Berezinskii-Kosterlitz-Thouless (BKT) behavior~\cite{berezinskii1971,kosterlitz1973,nelson1977,prokofev2001,prokofev2002}. The density is a smooth function of the chemical potential and shows weak finite-size effects, see Fig.~\ref{fig:results}(a1)~\cite{note:muprime}.
In constrast, the SF density shows strong size dependence, see Fig.~\ref{fig:results}(a2).
For low chemical potential, $\ns$ scales down with $L$, pointing towards a NF phase, while for high chemical potential, it converges to a finite value, as expected in the SF phase.
See also Fig.~\ref{fig:results}(a3), which shows the variation of $\ns$ with the system size for various values of the chemical potential.
This behaviour is consistent with the BKT universal jump at criticality, $\ns\lambdaT^2 = 4$ with $\lambdaT=\sqrt{2\pi\hbar^2/m\kB T}$ the thermal de Broglie wavelength.
It allows us to precisely locate the NF-to-SF transition point as the chemical potential $\muc'$ such that $\ns\lambdaT^2 = 4$ for the largest considered sizes. We use a conservative errorbar for the critical chemical potential corresponding to the variation of $\mu'$ with the system size in the range $L/a\in[20,60]$, see shaded area in Fig.~\ref{fig:results}(a2).
Although it can be refined using appropriate finite-size scaling~\cite{prokofev2001}, it appears to be sufficient for our purpose.
The corresponding critical density,  $\nc'$, is then found using the equation of state (particle density versus temperature and chemical potential) as found from QMC calculations, see Fig.~\ref{fig:results}(a1).
For the parameters of Fig.~\ref{fig:results}(a), it yields $\muc' = 0.052 \pm 0.004$ and  $\nc' = 0.29 \pm 0.03$.

We now turn to the behavior of the Bose gas in the presence of the quasicrystal potential.
Firstly, the NF regime is found by combining the above results with their counterparts at $V_0 \neq 0$ [Fig.~\ref{fig:results}(b), right column]. For a given interaction strength and temperature, we use the equation of state at $V_0 \neq 0$ to infer the chemical potential $\muc^1$ corresponding to the critical density of the homogeneous gas, $\nc^1=\nc'$, see Fig.~\ref{fig:results}(b1). It yields the NF-BG threshold shown on the phase diagrams of Fig.~\ref{fig:phasediagrams}.
Note that at $\muc^1$, we find a finite compressibility $\kappa =\partial n / \partial \mu$ [finite slope in Fig.~\ref{fig:results}(b1)] and a vanishingly small $\ns$ [see Fig.~\ref{fig:results}(b2)], which allows us to discriminate the BG against the SF and the MI.

Secondly, having identified the NF regime, we can focus on the BG-to-SF transition. Compared to the homogeneous case, the QMC results in the presence of the quasicrystal potential show stronger finite-size effects of both quantities $n$ and $\ns$.
The equation of state shown on Fig.~\ref{fig:results}(b1) is the density versus chemical potential averaged over the system size in the range $L \in [20,50]$
with the shaded area corresponding to the standard deviation.
On top of these fluctuations, the SF density nevertheless shows a clear finite-size scaling, qualitatively reminiscent of that found in the homogeneous gas at the NF-to-SF transition, see Fig.~\ref{fig:results}(b2).
The Inset of Fig.~\ref{fig:results}(b2) is a magnification in the vicinity of the transition with more system sizes where the fluctuations of $\ns$ versus $L$ are more clearly seen.
We find that the SF density sharply crosses over from vanishingly small values to a few units of $1/\lambdaT^2$.
We then locate the SF transition in the middle of the interval of chemical potentials such that $3 \leq \ns\lambda_T^2 \leq 5$ for all system sizes in the range $L/a\in[30,50]$, the errorbar corresponding to the size of this interval.
The BG-to-SF transition obtained here is clearly distinguished from the NF-BG threshold.
For instance, for the parameters of Fig.~\ref{fig:results}, we find $\muc^1 = 4.28 \pm 0.02$ and  $\nc^1 = 0.29 \pm 0.03$
at the NF-BG threshold and $\muc^2 = 4.39 \pm 0.02$ and  $\nc^2 = 0.55 \pm 0.05$ at the BG-to-SF transition.

The values of $\muc^1$ and $\muc^2$ versus $T$ hence obtained are used to locate the NF-BG threshold and the BG-to-SF transition on the phase diagrams of Fig.~\ref{fig:phasediagrams}, together with the corresponding errorbars.

\begin{figure}[t!]
         \centering
         \includegraphics[width = 0.95\columnwidth]{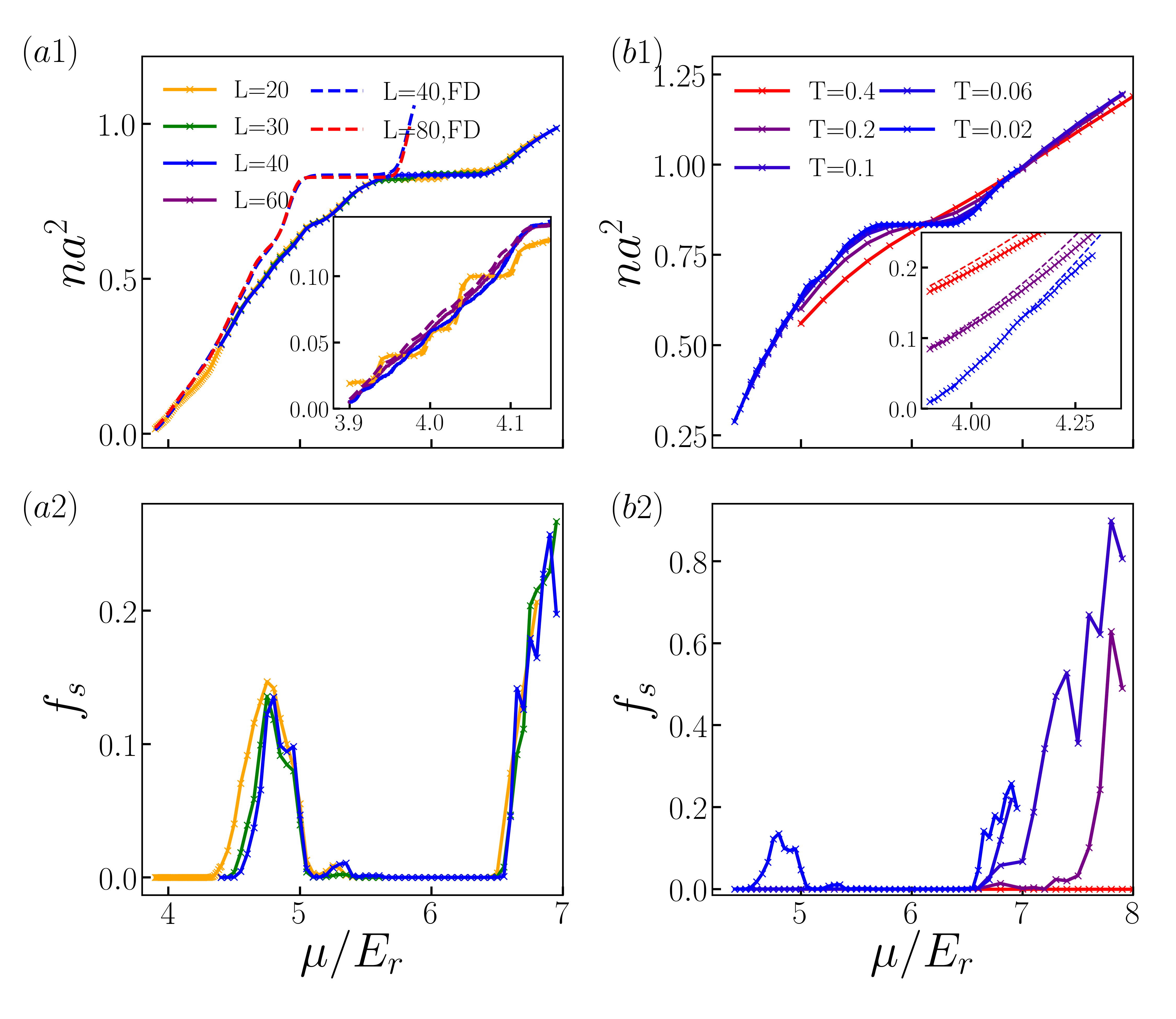}     
\caption{\label{fig:StronglyInteracting}
Strongly-interacting regime, $\tildegnot=5$.
The upper and lower rows show, respectively, the total particle density and the SF fraction versus the chemical potential.
Left column: Low-temperature regime, $T=0.02\Er/\kB$, for various system sizes.
The Inset of~(a1) shows the low-density regime for even lower temperature,  $T=0.0025\Er/\kB$.
Right column: Behaviour for various temperatures and a system size $L=40a$.
QMC results for the interacting Bose gas are shown as markers and solid lines,
while the Fermi-Dirac (FD) predictions, Eq.~(\ref{eq:FD}), are shown as dashed lines.
}
 \end{figure}

\lettersection{MI phase}\label{sec:MI}
We now turn to the strongly-interacting regime ($\tildegnot \gg 1$), where MI lobes emerge, see Fig.~\ref{fig:phasediagrams}(c).
Typical QMC results for the density and superfluid fraction are shown on Fig.~\ref{fig:StronglyInteracting},
for (a)~vanishingly small and (b)~finite temperatures.
The different line colors correspond to different sizes on panel~(a) and different temperatures on panel~(b).
For a weak chemical potential, the bosons populate the low-lying single-particle states, where strong repulsive interactions suppress multiple occupancy.
This mimics Pauli exclusion in real space and a simple fermionization picture accounts for the equation of state of the strongly-interacting Bose gas, within the Fermi-Dirac distribution,
\begin{equation}\label{eq:FD}
n \simeq \frac{1}{L^2}\sum_j \frac{1}{\e^{(E_j-\mu)/\kB T}+1},
\end{equation}
where $j$ spans the set of single particle states, with energy $E_j$.
This formula (dashed lines) indeed shows good agreement with the QMC results (solid lines) at vanishing, as well as finite temperatures and low chemical potential,
see Insets of Fig.~\ref{fig:StronglyInteracting}(a1) and (b1).

Consider first the low-temperature regime.
The lowest states are localized and, owing to the eightfold rotational symmetry of the quasicrystal potential,
they are arranged in rings of 8 or 16 trapping sites.
The interacting Bose gas then organizes in MI rings, characterized by Mott plateaus at commensurability, see Inset of Fig.~\ref{fig:StronglyInteracting}(a1).
Out of commensurability, finite tunneling between the trapping sites of a given ring generates ring superfluidity, but energy gaps between the different rings prevent long-range superfluidity, hence creating a BG phase.
We consistently find that the SF fraction vanishes for $\mu \lesssim 4.4\Er$, see Fig.~\ref{fig:StronglyInteracting}(a2).
Similar phenomenology was observed in small systems in Ref.~\cite{gautier2021}.
However, when the system size increases, new rings with slightly shifted energies appear.
This progressively fills the smallest gaps and blurs the corresponding Mott plateaus as observed in our QMC results when the system size increases,
see Inset of Fig.~\ref{fig:StronglyInteracting}(a1).
In the thermodynamic limit, the compressibility is thus finite and we find a BG.
In contrast, the QMC results show that the largest gaps survive when the system size increases, hence creating legitimate MI phases.
This occurs, for instance, for $\tildegnot=5$ and $5.5\Er \lesssim \mu \lesssim 6.4\Er$,
see Fig.~\ref{fig:phasediagrams}(c) as well as Figs.~\ref{fig:StronglyInteracting}(a1) and (b1).
This is consistent with the survival of a single-particle gap
and the existence of a plateau in the Fermi-Dirac prediction~(\ref{eq:FD}) at the same density and even larger systems, see Figs.~\ref{fig:StronglyInteracting}(a1).
Here, however, the chemical potential is high enough to populate many states, made of a large number of trapping sites, with nonzero spatial overlap. This generates a finite, positive interaction energy, which contributes to the chemical potential and correspondingly shifts the QMC results for interacting bosons compared to the Fermi-Dirac distribution.

We finally discuss the finite-temperature effects. 
When the temperature increases, the Mott plateaus shrink. The compressibility becomes finite but the SF fraction remains zero, hence progressively opening BG phases on the edges of the Mott plateaus, see Figs.~\ref{fig:StronglyInteracting}(b1) and (b2). For low enough temperature, the plateaus are still marked with very small compressibility and we identify 
$\kappa < 0.01 m/\hbar^{2} $ to a finite-temperature MI regime, corresponding to the MI lobes in the phase diagram of Fig.~\ref{fig:phasediagrams}(c).
As expected, finite temperatures also suppress the SF fraction in the SF phases and give space to the BG when it vanishes, see Fig.~\ref{fig:StronglyInteracting}(b2). Note that, here, the Bose gas is a superfluid in the absence of the quasicrystal potential, hence the compressible insulator we obtain is a legitimate finite-temperature BG.

\lettersection{Conclusion}
In conclusion, we have established the first thermodynamic phase diagrams of weakly to strongly interacting 2D bosons in a quasicrystalline potential.
The quantum phases are obtained analyzing finite-size effects and systematically distinguished from the NF regime.
Our results show the emergence of a sizable BG phase induced by the quasicrystalline potential.
For the parameters used here, the BG extends over a range where the density typically varies by a factor from $2$ to $4$ in all phase diagrams of Fig.~\ref{fig:phasediagrams}, where the considered temperatures are relevant for ultracold-atom experiments.
This paves the way to the direct observation of the BG in quantum simulators.
Moreover, further calculations not presented here show that the BG phase survives up to
$T \simeq 8\Er/\kB$ for $\tilde{g}_0 = 0.05$,
$T \simeq 0.7\Er/\kB$ for $\tilde{g}_0 = 0.86$,
and $T \simeq 0.5\Er/\kB$ for $\tilde{g}_0 = 5$.

Our results would directly apply to experiments performed in optical boxes~\cite{gaunt2013,chomaz2015,navon2021}.
For experiments performed in confining traps, our diagrams, found versus chemical potential, are amenable to local density approximation (LDA).
It applies provided the variation of the trap potential is negligible over a large enough distance such that the finite size effects become insignificant. Our results show that a size $L\sim 40a$ is a minimum.
For the parameters of Refs.~\cite{viebahn2019,sbroscia2020} for instance, it corresponds to a variation of $\simeq 0.01E_r$ from the trap center, smaller than the typical energy scales in our phase diagrams, and LDA is well applicable.

Moreover, our work raises new questions, notably about the nature of the SF-to-BG transition.
Our results are phenomenologically similar to a BKT transition, but the exact mechanism at the origin of the transition, as well as the effect of the quasicrystal potential on vortex pairing remain to be elucidated, via quantum simulation experiments and theoretical work.

\begin{acknowledgments}
We thank Dean Johnstone and Tommaso Macri for useful comments on the manuscript.
We acknowledge the CPHT computer team for valuable support.
This research was supported by the Paris region DIM-SIRTEQ, the IPParis Doctoral School and the Swiss National Science Foundation under Division II.
This work was performed using HPC/AI resources from GENCI-CINES and GENCI-TGCC (Grants 2020-A0090510300 and 2021-A0110510300),
and made use of the ALPS scheduler library and statistical analysis tools~\cite{troyer1998,ALPS2007,ALPS2011}.
\end{acknowledgments}


\begin{thebibliography}{10}
\bibitem{note:muprime}
{To ease comparison to the quasicrystal case, the chemical potential is noted
  with a prime for the homogeneous case.}
\end{thebibliography}

\end{document}